\documentclass[prl,twocolumn,
showpacs,amsmath,amssymb]{revtex4}
\usepackage[dvips]{graphicx}
\usepackage{enumerate}
\newcommand \beq{\begin{eqnarray}}
\newcommand \eeq{\end{eqnarray}}

\begin{document}

\title{Dense QCD in a Finite Volume}

\author{Naoki Yamamoto and Takuya Kanazawa} 
\affiliation{Department of Physics, The University of Tokyo, Tokyo 113-0033, Japan\\}

\begin{abstract}

We study the properties of QCD at high baryon density 
in a finite volume where color superconductivity occurs.
We derive exact sum rules for complex eigenvalues of 
the Dirac operator at finite chemical potential, 
and show that the Dirac spectrum is directly related to 
the color superconducting gap $\Delta$.
Also, we find a characteristic signature of color superconductivity:
an X-shaped spectrum of partition function zeros in the complex quark mass plane 
near the origin, 
reflecting the $Z(2)_{L} \times Z(2)_{R}$ symmetry of the diquark pairing.
Our results are universal in the domain 
$\Delta^{-1} \ll L \ll m_{\pi}^{-1}$ where $L$ is the linear size of the system
and $m_{\pi}$ is the pion mass at high density.

\end{abstract}


\maketitle

%{\it Introduction}\ ---
Revealing Quantum Chromodynamics (QCD) in the regime of finite
temperature ($T$) and chemical potential ($\mu$) is important for
understanding a wide range of phenomena from ultrarelativistic heavy
ion collisions, the early Universe, and neutron stars to possible 
quark stars \cite{YHM}.
A lot of theoretical progress has been made by the first-principles
lattice QCD Monte Carlo simulations 
in the study of the finite-$T$ regime \cite{K}.
However, the application of the lattice technique to QCD at 
finite $\mu$ is still hampered by the notorious fermion sign problem: 
calculation of the QCD partition function  
requires dealing with a path integral with a measure including
a complex fermion determinant.
This is the main reason why our understanding of the 
properties of QCD at finite $\mu$ is still immature,
except at asymptotic high $\mu$ where 
the ground state is shown to be
the most symmetric three-flavor ($N_f=3$) color superconductivity (CSC),
i.e., the color-flavor locked (CFL) phase \cite{CSC, ARW}
by using the weak QCD coupling calculations.

In this Letter, we demonstrate exact analytical results for QCD at 
high $\mu$ specific for a large but finite volume.
By matching the partition function of QCD
against that of the effective theory of CSC, we derive exact
sum rules for the Dirac eigenvalues (Dirac spectrum)
as well as the spectrum of partition function zeros 
(Lee-Yang zeros \cite{YL}) in the complex quark mass $m$-plane.
As is well known at $\mu=0$, 
the Dirac spectrum and the Lee-Yang zeros spectrum 
of QCD in a finite volume are very closely
related to the chiral symmetry breaking:
the chiral condensate $\langle \bar q q \rangle$
is directly connected to the Dirac spectrum by the exact relations 
such as the Banks-Casher relation \cite{BC80} and
the Leutwyler-Smilga sum rules \cite{LS92}.
Also, a nonzero $\langle \bar q q \rangle$
implies the existence of a line of the Lee-Yang zeros going 
through $m=0$ in the complex $m$-plane 
independent of $\mu$ \cite{note}.
Nevertheless, such exact relations at finite $\mu$ 
and the relevance of the Lee-Yang zeros spectrum to the CSC 
have not been fully understood.

As we shall show below, the Dirac spectrum at high $\mu$ 
is intimately related to the CSC gap $\Delta$, 
rather than to $\langle \bar q q \rangle$,
through our spectral sum rules. Also, a nonzero gap $\Delta$
necessitates an X-shaped cut of Lee-Yang zeros near $m=0$.
In particular, the $Z(2)_L \times Z(2)_R$ symmetry of the diquark pairing 
plays a crucial role on both Dirac and Lee-Yang zeros spectra.
Together with the exact results at $\mu=0$ \cite{LS92}, 
we expect that our results impose strong constraints on their possible spectra 
and provide important insights to the properties of QCD at finite $\mu$.

In the following, we will focus on QCD with $N_f=3$ 
(light up, down and strange quarks)
at finite quark chemical potential $\mu$
living on the four-dimensional torus
$V_4=L \times L \times L \times \beta$ with $\beta=1/T \sim L$.
%{\it QCD at high density}\ ---
Let us consider the Euclidean QCD Lagrangian
defined as ${\cal L_{\rm QCD}}= \bar q ( {\hat {\cal D}} + {M}) q + {\cal L}_{g}$ with
${\cal L}_{g}=\frac{1}{4}F_{\mu \nu}F^{\mu \nu}$ and the Dirac operator
${\hat {\cal D}}=\gamma^{\mu}(\partial_{\mu}+ig A_{\mu}) + \mu\gamma_0$.
Here $q$ is the quark field and 
$A_{\mu}=A^a_{\mu}t^a$ is the gluon field with the color $SU(3)_C$ 
generators $t^a$ $(a=1, 2, \cdots, 8)$.
$M$ is the complex quark mass matrix, $g$ is the QCD
coupling constant and 
$F_{\mu \nu}=\partial_{\mu}A_{\nu} -\partial_{\nu}A_{\mu}+ig[A_{\mu}, A_{\nu}]$.
Since ${\hat {\cal D}}$ is not anti-Hermite with $\mu>0$, 
its eigenvalues $i\lambda_n$ are generally complex values, 
whereas $i\lambda_n$ are pure imaginary at $\mu=0$.
Even so, ${\hat {\cal D}}$ preserves the chirality,
$\{\gamma_5, {\hat {\cal D}}\}=0$. 
The chirality ensures that if $i\lambda_n$ is the eigenvalue of 
${\hat {\cal D}}$ (${\hat {\cal D}}\psi_n=i\lambda_n \psi_n$),
then so is $-i\lambda_n$.

The QCD partition function $Z_{\rm QCD}$ involves 
a sum over the different topological sectors of the gauge-field configurations 
characterized by the integer topological charge $\nu$ as 
$Z_{\rm QCD}=\sum_{\nu}
e^{i \nu \theta} Z_{\nu}$.
At $\mu \gg \Lambda_{\rm QCD}$ ($\Lambda_{\rm QCD}$: the typical scale of QCD), 
however, the topological susceptibility is highly suppressed as 
$\langle \nu^2 \rangle \propto (\Lambda_{\rm QCD}/\mu)^{8}$ \cite{S02}
owing to the screening of instantons in the medium together 
with the asymptotic freedom of QCD.
Thus, we can focus on the topological sector $\nu=0$ alone.

The QCD partition function with $\nu=0$ can be written in the functional integral
using the symmetry $i\lambda_n \leftrightarrow -i\lambda_n$:
\beq
\label{eq:part_QCD}
Z_{\rm QCD}
=
\left\langle \! \! \! \left\langle 
\prod_{{\rm Re}(\lambda_n)>0} 
\det  \left(1 +  \frac{M^{\dag}M}{\lambda_n^2} \right)  
\right\rangle \! \! \! \right\rangle,
\eeq
where
$\langle \! \langle {\cal O} \rangle \! \rangle
={\int [dA]  {\cal O}  e^{-S_{g}} 
( \prod_n \lambda _n^2 
)^{N_f}  }/
{\int [dA] e^{-S_{g}} 
( \prod_n \lambda _n^2 
)^{N_f}}$
is the average of ${\cal O}$ over all gauge configurations
with $S_g$ being the classical action of the gluon field.
$Z_{\rm QCD}$ is normalized so that $Z_{\rm QCD}=1$ when quark
masses are turned off.

%{\it Effective theory of color superconductivity}\ ---
We shall give the partition function $Z_{\rm EFT}$ 
from the effective theory of the color superconductivity (CSC). 
For definiteness, we consider the most predominant diquark pairing, 
the color-flavor locked (CFL) phase \cite{ARW}:
$\langle(q_L)^j_b C (q_L)^k_c \rangle 
\sim \epsilon_{abc} \epsilon_{ijk}[d_L^{\dag}]_{ai}$
and
$\langle(q_R)^j_b C (q_R)^k_c \rangle 
\sim \epsilon_{abc} \epsilon_{ijk}[d_R^{\dag}]_{ai}$
where $i,j,k$ ($a,b,c$) are the flavor (color) indices,
and $C$ is the charge conjugation matrix.
The remarkable feature here is that chiral symmetry is dynamically 
broken by the {\it diquark condensate}:
the symmetry breaking pattern of the CFL phase
at asymptotic high $\mu$ is
$SU(3)_C \times SU(3)_L \times SU(3)_R \times U(1)_B \times U(1)_A
\rightarrow SU(3)_{C+L+R} \times Z(2)_L \times Z(2)_R$
[The $Z(2)_L \times Z(2)_R$ symmetry left reflects the fact that 
we can change the sign of the left-handed or 
right-handed quark fields independently].
As a result, we have $8+1+1$ Nambu-Goldstone (NG) modes
associated with the breaking of chiral symmetry, $U(1)_A$ and $U(1)_B$
symmetries, which we will refer to as pions, $\eta'$, and $H$, respectively.
In the following, 
we will not consider $H$ since its dynamics decouples.
In the CFL phase, we also have gluons and quarks; 
the gluons acquire a mass comparable to the CSC gap $\Delta \propto |d_{L}|/\mu^2$
\cite{CGN01, HTY08} by the Anderson-Higgs mechanism 
when the $SU(3)_C$ symmetry is broken;
the octet (singlet) quarks of the unbroken $SU(3)_{C+L+R}$ symmetry
have the mass gap $\Delta$ ($2\Delta$) \cite{ARW}. 

We then specify the microscopic domain (or $\epsilon$-domain) of the CFL phase.
The microscopic domain of QCD at $\mu=0$ is specified by 
$\Lambda^{-1} \ll L \ll m_{\Pi}^{-1}$,
where $m_{\Pi}$ is the mass of pions at $\mu=0$
%associated with chiral symmetry breaking at low $\mu$ 
and $\Lambda$ is the mass scale of the lightest non-NG modes 
(i.e., the $\rho$ meson mass $m_{\rho}$) \cite{LS92}.
The corresponding microscopic domain of the CFL can be defined as
\beq
\label{eq:regime}
\frac{1}{\Delta} \ll L \ll \frac{1}{m_{\pi}},
\eeq
where $m_{\pi}$ is the mass of pions associated with the CSC at high $\mu$. 
The first condition in Eq. (\ref{eq:regime}) follows by comparing
the contribution to $Z_{\rm EFT}$ of the pions, $e^{-m_{\pi}L}$, to that
of the other heavier particles, $e^{-\Delta L}$. 
This condition allows us only to deal with the pions 
described by the CFL effective Lagrangian.
On the other hand, the second condition in Eq. (\ref{eq:regime}) means that the
Compton wavelength of the pions is much larger than the linear size of the box,
so that the CFL effective Lagrangian can be truncated to its zero momentum sector.
Note that the second condition is automatically satisfied at sufficiently high $\mu$
with $L$ and quark mass $m$ fixed,
since $m_{\pi} \sim \Delta m/\mu$ 
(see Eq. (\ref{eq:EFT}) below)
together with the relation 
$\Delta \sim \mu \exp\left(-\frac{3\pi^2}{\sqrt{2}g} \right)$ \cite{S99}. 
Note also that, in the domain (\ref{eq:regime}), temperature $T$
is low enough for the CSC to be realized
since $T \ll \Delta \sim T_c$ with
$T_c$ being the critical temperature of the CSC.

The CFL effective Lagrangian in the Minkowski space-time
up to the leading order ${\cal O}(M^2)$ is given by \cite{CG99}
\beq
\label{eq:EFT}
 {\cal L_{\rm EFT}}
&=& \frac{f_{\pi}^2}{4}{\rm Tr}
[\nabla_0 \tilde\Sigma \nabla_0 \tilde\Sigma^{\dag} - 
v_{\pi}^2 \partial_i \tilde\Sigma \partial \tilde\Sigma^{\dag}]
\nonumber \\
&+& \frac{3f_{\eta'}^2}{4} \left[\partial_0 V \partial_0 V^{*} 
 -  v_{\eta'}^2 \partial_i V \partial_i V^{*} \right]
 \\
&+&
\frac{3 \Delta^2}{4\pi^2} \left[V({\rm Tr} M \tilde\Sigma^{\dag})^2
 - V{\rm Tr} (M \tilde\Sigma^{\dag})^2 + {\rm H.c.} \right],
\nonumber
\eeq
where $\tilde\Sigma=\exp(i\pi^a \lambda^a/f_{\pi})$ and 
$V=\exp[2i\eta'/(\sqrt{\mathstrut 6} f_{\eta'})]$ 
are the pion and $\eta'$ fields respectively, $f_{\pi}\ (f_{\eta'})$ is the pion ($\eta'$)
decay constant, $v_{\pi}\ (v_{\eta'})$ is the pion ($\eta'$) velocity,
$\lambda^a$ $(a=1,2, \cdots, 8)$ are the Gell-Mann matrices,
and the covariant derivative including the effective chemical potential
(Bedaque-Sch\"{a}fer term \cite{BS02}) is given by
$\nabla_0 \tilde\Sigma
=\partial_0 \tilde\Sigma + i\left( \frac{MM^{\dag}}{2p_F} \right) \tilde\Sigma 
- i\tilde\Sigma \left( \frac{M^{\dag}M}{2p_F} \right)$
with the Fermi momentum $p_F$.
The quantities $f_{\pi}$ and $f_{\eta'}$ 
can be perturbatively computed at $\mu \gg \Lambda_{\rm QCD}$ as 
$\frac{f_{\pi}^2}{p_F^2}=\frac{21-8\log2}{36 \pi^2}$ and 
$\frac{f_{\eta'}^2}{p_F^2}=\frac{3}{8 \pi^2}$ \cite{CG99}.

In Eq. (\ref{eq:EFT}), we have neglected the mass term of order ${\cal O}(M)$, 
since this term originates from the instanton 
contribution and is suppressed at asymptotic high $\mu$ \cite{S02}.
Thus, the leading mass term in the CFL effective Lagrangian is ${\cal O}(M^2)$, 
unlike the ${\cal O}(M)$ term in the usual chiral Lagrangian at low $\mu$. 
A more intuitive explanation for this fact is that 
$M \bar q q$ is prohibited by the $Z(2)_L \times Z(2)_R$ symmetry,
but $(M \bar q q)^2$ is not.

In the domain $L \ll m_{\pi}^{-1}$, one can neglect the contribution 
of the kinetic term.
Then the partition function for the CFL effective Lagrangian reads:
\beq
\label{eq:part_EFT}
Z_{\rm EFT}\! \! \! &=&\! \! \! \! \int \! \! d\Sigma 
\! \exp \! \left(\! V_4\frac{3\Delta ^2}{4\pi ^2}
\! \left[ ({\rm Tr} M\Sigma^{\dag} )^2 
\! - \! {\rm Tr}(M\Sigma^{\dag})^2 \right]{\rm det}\Sigma 
\! + \! {\rm  H.c.}
\right),
\nonumber \\
\eeq
where the integral is over $\Sigma \equiv \tilde\Sigma V \in U(3)$ and 
$Z_{\rm EFT}$ is normalized so that $Z_{\rm EFT}=Z_{\rm QCD}=1$ 
in the chiral limit.
In Eq. (\ref{eq:part_EFT}), we have also neglected 
the effect of the effective chemical potential.
[If one includes it, Eq. (\ref{eq:part_EFT}) can be expanded in terms of 
not only $(V\Delta^2)^2{\cal O}(M^4)$ but also $V {\cal O}(M^4)$. 
In the domain $\Delta^{-1} \ll L$, however, the latter is negligible.]

Owing to the property of $\Sigma \in U(3)$, 
$[({\rm Tr} M\Sigma^{\dag})^2 - {\rm Tr}(M\Sigma^{\dag})^2]{\rm det}\Sigma
=2{\rm det}M {\rm Tr}(M^{-1}\Sigma)$,
Eq. (\ref{eq:part_EFT}) can be expressed analytically 
as shown in Ref. \cite{JSV96}.
In particular, in the flavor symmetric case $M=m{\bf 1}$,
one can evaluate $Z_{\rm EFT}$ in a simpler form 
using Weyl's formula \cite{BG79, LS92}:
\beq
\label{eq:part_EFT3}
Z_{\rm EFT}=
\mathop {\det }\limits_{0 \le i,j \le 2} \left[I_{j- i} (x)\right],\label{eq:det}
\eeq
where $I_{\nu}(x)$ is the modified Bessel function and $x=3V_4 m^2 \Delta^2/\pi^2$.
The combination of $m^2 \Delta^2$ is expected,
since $m^2$ acts as a source for $\Delta^2$.
It should be remarked that the expression (\ref{eq:part_EFT3}) is exactly the same form as 
the partition function $Z_0$ with $\nu=0$ at $\mu=0$
which is given by Eq. (\ref{eq:part_EFT3}) 
with the replacement of the argument: $x \rightarrow x'= V_4 m |\langle \bar q q \rangle|$.
This is a novel correspondence 
between the CSC phase and the hadronic phase,
and may have relevance to the idea of 
their continuity \cite{SW99}.
In Table.~\ref{tab}, we summarize our main results at $\mu \gg \Lambda_{\rm QCD}$ below
%(which we will obtain below)
compared with the results at $\mu=0$ in Ref.~\cite{LS92}.

%{\it Spectral sum rules}\ ---
Expanding in terms of quark mass $M$
and performing the group integral over $\Sigma \in U(3)$ order by order,
Eq. (\ref{eq:part_EFT}) reduces to the following form up to ${\cal O}(M^6)$:
\beq
\label{eq:part_EFT2}
\! \! Z_{\rm EFT} \!\sim\! 1+\frac{3}{8}\left( \! V_4\frac{\Delta^2}{\pi^2} \! \right)^{\! \! 2} \! \!
\left[({\rm Tr}M^{\dag}M)^2 \!-\! {\rm Tr}(M^{\dag}M M^{\dag}M)  \right]\! \!.
\eeq
Using the relation,
$\det[1+\epsilon] = 1 + {\rm Tr} \epsilon 
+ \frac{1}{2}[{({\rm Tr}\epsilon)^2 - {\rm Tr}\epsilon^2}]
+ {\cal O}(\epsilon^3)$,
one can expand the QCD partition function (\ref{eq:part_QCD}) in terms of the
quark mass matrix $M$.
Then one obtains the spectral sum rules for the Dirac eigenvalues $i\lambda_n$
by matching this expansion against Eq. (\ref{eq:part_EFT2}). 
By rescaling $z_n=\sqrt{V_4}\Delta \lambda_n$, the results read 
\beq
\label{eq:sr1}
\left\langle \! \! \! \left\langle 
 {\sum \limits_{n}}' \frac{1}{z_n^4}  \right\rangle \! \! \! \right\rangle 
\! &=& \! 
\left\langle \! \! \! \left\langle 
\left( {\sum \limits_{n}}' \frac{1}{z_n^2} \right)^{\! \! 2} 
\right\rangle \! \! \! \right\rangle
\! = \! \frac{3}{4\pi^4},
\\
\label{eq:sr2}
\left\langle \! \! \! \left\langle 
 {\sum \limits_{n}}'  \frac{1}{z_n^2}  \right\rangle \! \! \! \right\rangle
\! &=& \! 
\left\langle \! \! \! \left\langle 
 {\sum \limits_{n}}'  \frac{1}{z_n^6}  \right\rangle \! \! \! \right\rangle
\! = \!
\left\langle \! \! \! \left\langle 
\left({\sum \limits_{n}}' \frac{1}{z_n^2}\right)^{\! \! 3} 
\right\rangle \! \! \! \right\rangle
\nonumber \\ 
\!  &=&  \! \left\langle \! \! \! \left\langle \!
\left({\sum \limits_{n}}' \frac{1}{z_n^2}\right) \! \!
\left({\sum \limits_{n}}'  \frac{1}{z_n^4}\right) \!
 \right\rangle \! \! \! \right\rangle
\! = 0,
\eeq
where the summation $\sum'$ is taken over $z_n$
satisfying ${\rm Re}(z_n)>0$ and 
$|z_n| \lesssim \sqrt{V_4}\Delta^2$ ($|\lambda_n|\lesssim \Delta$).
These relations are highly nontrivial,
since sums of inverse powers of {\it complex} $\lambda_n$
with the average taken over the gauge configurations
give the {\it real} value involved with the CSC gap $\Delta$.
In particular, the sums in Eq. (\ref{eq:sr2}) are identically zero, 
which is a direct consequence of the $Z(2)_L \times Z(2)_R$ symmetry 
of the quarks at high $\mu$.
This situation should be compared with QCD at $\mu=0$ \cite{LS92}
(at low $\mu$ \cite{A07}),
where sums of inverse powers of {\it real} $\lambda_n$ 
({\it complex} $\lambda_n$)
take {\it positive} values involved with 
the chiral condensate $\langle \bar q q \rangle$.
As in the case of $\mu=0$ \cite{SV93,VZ93},
our spectral sum rules must be universal
(i.e., independent of microscopic details)
in the domain (\ref{eq:regime}).

By using the spectral density
$\rho(\lambda)=\langle \! \langle 
 \sum_n \delta^2(\lambda-\lambda_n) \rangle \! \rangle$,
the first sum rule in Eq. (\ref{eq:sr1})
reduces to the relation: $\mathop
\int_{{\mathbb C}_{+}} \frac{d^2 z}{z^4}
\frac{1}{V_4 \Delta^2}\rho\left(\frac{z}{\sqrt{V_4}\Delta}\right)
=\frac{3}{4\pi^4}$,
where ${\mathbb C}_+ =\{z: {\rm Re}(z)>0 \}$ and
$d^2 z={d ({\rm Re}z)}{d ({\rm Im}z)}$.
This implies the existence of the microscopic limit of
the spectral density defined as
\beq
\label{eq:micro}
\rho_s(z)=\mathop {\lim }\limits_{V_4 \to \infty } 
\frac{1}{V_4 \Delta^2}\rho\left(\frac{z}{\sqrt{V_4}\Delta}\right).
\eeq
From Eq. (\ref{eq:micro}), we find that 
the microscopic spectral density at high $\mu$
is governed by the CSC gap $\Delta$.
Also Eq. (\ref{eq:micro}) shows that 
the linear spacing of eigenvalues $\delta \lambda_n$
in the complex $\lambda$-plane satisfies 
$\delta \lambda_n \propto {1}/{\sqrt{V_4}}$.
Since $\delta \lambda_n \propto 1/V_4$ 
at low $\mu$ with $\langle \bar qq \rangle \neq 0$ \cite{LS92},
and $\delta\lambda_n \propto 1/V_4^{1/4}$ in a free theory, 
our result indicates a sizable deformation of the Dirac spectrum 
due to the dynamics of the CSC. 
We expect that the random matrix theory (RMT) \cite{VZ93} 
or the supersymmetric approach \cite{OTV99}
incorporating the CSC and the symmetries of the CFL
not only reproduce the above results, but also clarify the concrete form of 
the microscopic spectral density $\rho_s$.

\begin{table}[t]
\begin{tabular}{|c|c|c|c|c|}
\hline 
        & order parameter & source & $\delta \lambda_n$ & Lee-Yang zeros \\ \hline \hline
$\mu \gg \Lambda_{\rm QCD}$ & $\Delta^2 \propto |d_L^{\dag} d_R|/\mu^4$ 
& $m^2$ & $\propto 1/\sqrt{V_4}$ & ${\rm Re}(m)=\pm{\rm Im}(m)$ \\ \hline
$ \mu=0$ & $\langle \bar q_L q_R \rangle$ & $m$ & $\propto 1/V_4$ & ${\rm Re}(m)=0$ \\  \hline
\end{tabular}
\caption{The summary of our main results at $\mu \gg \Lambda_{\rm QCD}$
compared with the results at $\mu=0$ in Ref. \cite{LS92}:
order parameters, the corresponding source terms, 
the linear spacing of eigenvalues 
$\delta \lambda_n$, and the spectra of partition function zeros 
(Lee-Yang zeros) in the complex $m$-plane, respectively.}
\label{tab}
\end{table}

%{\it Partition function zeros}\ ---
Let us consider the partition function zeros (Lee-Yang zeros)
in the complex $m$-plane in the flavor-symmetric case.
Using the asymptotic form of $I_{\nu}(x)$, 
we find the partition function (\ref{eq:part_EFT3}) for $|x| \gg1$ as
\beq 
Z_{\rm EFT}(x=-iz) \sim z^{-5/2}\cos\left(4z-\frac{3\pi}{4} \right).
\eeq 
Then, the Lee-Yang zeros for $|x| \gg 1$ are given by
$x=-i(n+\frac{1}{4})\frac{\pi}{4}$ ($n \in {\mathbb Z}$).
Remembering $x=3V_4 m^2 \Delta^2/\pi^2$, 
the zeros are spaced 
along the lines 
\beq
\label{eq:LY}
{\rm Re}(m)=\pm {\rm Im}(m),
\eeq 
at $\mu \gg \Lambda_{\rm QCD}$. 
In the thermodynamic limit $V_4 \rightarrow \infty$, 
the density of the zeros increases and they
join into a cut in the vicinity of massless limit $m=0$.
However, the cut does not go through $m=0$, since
$dx/dm = 6V_4 m \Delta^2/\pi^2$ and
the density of the zeros vanishes at $m = 0$.

\begin{figure}[h]
\begin{center}
\includegraphics[width=7.0cm]{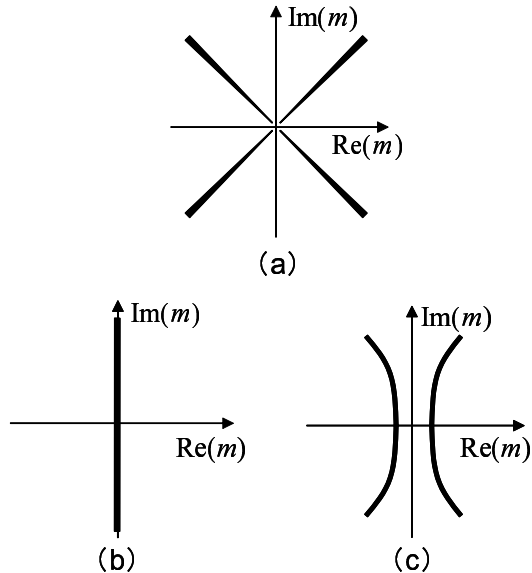}
\end{center}
\vspace{-0.6cm}
\caption{The spectra of the Lee-Yang zeros
in the complex $m$-plane in the vicinity of $m=0$
in the thermodynamic limit $V_4 \rightarrow \infty$:
(a) $\mu \gg \Lambda_{\rm QCD}$ (exactly obtained in our analysis),
(b) $\mu=0$ \cite{ALS02}, and
(c) $\mu>\mu_c$
with the critical $\mu_c$ of chiral symmetry restoration 
obtained from the random matrix theory \cite{HJV97}.
The density of the zeros fades away as $m \rightarrow 0$ in case (a),
while that remains constant in case (b).
}
\label{fig:LY}
\end{figure}

In Fig.~\ref{fig:LY},
we draw spectra of the Lee-Yang zeros 
in the complex $m$-plane near $m=0$:
the spectrum (a) at $\mu \gg \Lambda_{\rm QCD}$
is the {\it exact} result obtained above.
For comparison, we show
the spectra (b) at $\mu=0$ previously obtained
exactly \cite{ALS02},
and (c) for $\mu>\mu_c$ obtained from the RMT \cite{HJV97},
where $\mu_c$ is the critical chemical potential 
of chiral symmetry restoration.
In case (b), the density of the zeros at $m=0$ 
is finite as $dx'/dm=V_4 |\langle \bar q q \rangle|$,
and the chiral condensate $\langle \bar q q \rangle$ takes nonzero value.
As $\mu$ increases, the result (c) shows that 
the zeros will move away from the origin
and chiral symmetry restores \cite{HJV97}.

Our exact result, however, demonstrates that the scenario (c) of the RMT
should suffer from dramatic modifications 
if the effects of the CSC is taken into account:
even at $\mu \gg \Lambda_{\rm QCD}$, there exists an X-shaped cut
(\ref{eq:LY}) near $m=0$ as shown in Fig.~\ref{fig:LY}(a),
reflecting the $Z(2)_L \times Z(2)_R$ symmetry of the diquark pairing.
The distinctive feature of the cut (\ref{eq:LY}) is that there is a discontinuity
in the $m^2$-plane at $m=0$ along the real axis but not in the $m$-plane,
which is related to the fact that 
$\Delta \neq 0$ and $\langle \bar q q \rangle = 0$ at $\mu \gg \Lambda_{\rm QCD}$.
How the spectrum of the Lee-Yang zeros evolve from (b) to (a) 
as $\mu$ increases is beyond the scope of this paper 
and needs further investigations.
Note that summing over the topological sector $\nu$ 
would also change the scenario (c)
where $\nu=0$ alone is considered.

%-------- Figure 1-----------------------------
%----------------------------------------------

It is important to generalize our spectral sum rules or
to directly investigate the distributions of the partition function 
zeros at lower baryon densities. 
One can, e.g., match the QCD partition function at finite density 
against the effective theory of the generalized pions \cite{HTYB06}
in the entire span of the density 
where the microscopic regime can be defined as 
$m_{\rho}^{-1}\ll L \ll m_{\pi}^{-1}$ according to Ref.~\cite{HTY08}.
Also the generalization of our spectral sum rules to QCD-like theories,
such as the two-color QCD at high density,
would be an interesting problem to be investigated \cite{KWY}, 
which can be tested on the lattice QCD simulation.

We would like to thank T. Hatsuda for discussions, comments and reading the manuscript.
Discussions with S. Aoki, S. Sasaki and T. Wettig are greatly appreciated.
Author N. Y. is supported by the Japan Society for the Promotion of Science
for Young Scientists. 
Author T.K. is supported by Global COE Program 
``the Physical Sciences Frontier'', MEXT, Japan.

%%%%%%%%%%   REFERENCES   %%%%%%%%%%


\begin{thebibliography}{99}

\bibitem{YHM} Reviewed in, 
 K. Yagi, T. Hatsuda and Y. Miake, {\em Quark-Gluon Plasma},
 Cambridge Univ. press (Cambridge, 2005).

\bibitem{K} Reviewed in,
F. Karsch, Proc, Sci., CPOD07 (2007) 026; 
Proc. Sci., LAT2007 (2007) 015.

\bibitem{CSC}  Reviewed in, 
M. G. Alford, A. Schmitt, K. Rajagopal and T. Sch\"{a}fer, 
Rev. Mod. Phys. {\bf 80}, 1455 (2008).

\bibitem{ARW} 
M. G. Alford, K. Rajagopal and F. Wilczek, 
Nucl. Phys.{\bf B537}, 443 (1999).

\bibitem{YL}
C. N. Yang and T. D. Lee, 
Phys. Rev. {\bf 87}, 404 (1952);
T. D. Lee and C. N. Yang,
Phys. Rev. {\bf 87}, 410 (1952).

\bibitem{BC80}
T. Banks and A. Casher,
Nucl. Phys. {\bf B169}, 103 (1980). 

\bibitem{LS92}
H. Leutwyler and A. V. Smilga,
Phys. Rev. D {\bf 46}, 5607 (1992).

\bibitem{note}
When one factorizes the QCD partition function as
$Z(m) = \prod_n (m-m_n)$ ($m_n$: Lee-Yang zeros),
the chiral condensate is given by 
$\langle \bar q q \rangle 
=({1}/{V_4}) \sum_n {1}/{(m-m_n)}$. 
Then a nonzero chiral condensate in the thermodynamic limit
$V_4 \rightarrow \infty$ implies the convergence to $m=0$ of 
Lee-Yang zeros with an equidistant spacing $\sim 1/V_4$ \cite{HJV97}.

\bibitem{HJV97}
M. A. Halasz, A. D. Jackson, and J. J. M. Verbaarschot,
Phys. Rev. D {\bf 56}, 5140 (1997);
M. A. Halasz {\it et al.}, 
Phys. Rev. D {\bf 58}, 096007 (1998).

\bibitem{S02}
T. Sch\"{a}fer, Phys. Rev. D {\bf 65}, 094033 (2002);
N. Yamamoto, J. High Energy Phys. 12 (2008) 060. 

\bibitem{CGN01}
R. Casalbuoni, R. Gatto and G. Nardulli, 
Phys. Lett. B {\bf 498}, 179 (2001); 
V. P. Gusynin and I. A. Shovkovy,
Nucl. Phys. {\bf A700}, 577 (2002); 
H. Malekzadeh and D. H. Rischke, 
Phys. Rev. D {\bf 73}, 114006 (2006).

\bibitem{HTY08}
T. Hatsuda, M. Tachibana and N. Yamamoto,
Phys. Rev. D {\bf 78}, 011501 (2008). 

\bibitem{S99}
D. T. Son, 
Phys. Rev. D {\bf 59}, 094019 (1999). 

\bibitem{CG99}
R. Casalbuoni and R. Gatto,
Phys. Lett. B {\bf 464}, 111 (1999);
D.~T.~Son and M.~A.~Stephanov, 
Phys.\ Rev.\ D {\bf 61}, 074012 (2000);
{\bf 62}, 059902(E) (2000);
T. Sch\"{a}fer, Phys. Rev. D {\bf 65} 074006 (2002). 

\bibitem{BS02}
P. F. Bedaque and T. Sch\"{a}fer, 
Nucl. Phys. {\bf A697}, 802 (2002).

\bibitem{JSV96}
A. D. Jackson, M. K. Sener and J. J. M. Verbaarschot,
Phys. Lett. B {\bf 387}, 355 (1996). 

\bibitem{BG79}
I. Bars and F. Green, Phys. Rev. D {\bf 20}, 3311 (1979). 

\bibitem{SW99}
T. Sch\"{a}fer and F. Wilczek,
Phys. Rev. Lett. {\bf 82}, 3956 (1999). 

\bibitem{A07}
G. Akemann, Int. J. Mod. Phys. A {\bf 22}, 1077 (2007);
J. C. Osborn, K. Splittorff and J. J. M. Verbaarschot,
Phys. Rev. D {\bf 78}, 065029 (2008).

\bibitem{SV93}
E. V. Shuryak and J. J. M. Verbaarschot, 
Nucl. Phys. {\bf A560}, 306 (1993).

\bibitem{VZ93}
J. J. M. Verbaarschot and I. Zahed, 
Phys. Rev. Lett. {\bf 70}, 3852 (1993).

\bibitem{OTV99}
J. C. Osborn, D. Toublan and J. J. M. Verbaarschot,
Nucl. Phys. {\bf B540}, 317 (1999);
P. H. Damgaard {\it et al.}, 
Nucl. Phys. {\bf B547}, 305 (1999). 

\bibitem{ALS02}
G. Akemann, J. T. Lenaghan and K. Splittorff,
Phys. Rev. D {\bf 65}, 085015 (2002). 

\bibitem{HTYB06}
T. Hatsuda, M. Tachibana, N. Yamamoto and G. Baym, 
Phys. Rev. Lett. {\bf 97}, 122001 (2006);
N. Yamamoto {\it et al.}, 
Phys. Rev. D {\bf 76}, 074001 (2007).

\bibitem{KWY}
T. Kanazawa, T. Wettig and N. Yamamoto,
arXiv:0906.3579 [hep-ph].

\end{thebibliography}
\end{document}